\documentclass[doublecol]{epl2}

\title{Dynamical Inequality in Growth Models}

\author{Eytan Katzav\inst{1}\thanks{E-mail: \email{eytan.katzav@kcl.ac.uk}} \and Moshe Schwartz\inst{2}\thanks{E-mail: \email{bricki@netvision.net.il}}}
\shortauthor{E. Katzav \and M. Schwartz}
\institute{
  \inst{1} Department of Mathematics, King's College London, Strand, London WC2R 2LS, UK, EU\\
  \inst{2} Department of Physics, Raymond and Beverly Sackler Faculty of Exact Sciences, Tel Aviv University, Tel Aviv 69978, Israel
}
\pacs{64.60.Ht}{Dynamic critical phenomena}
\pacs{02.50.-r}{Probability theory, stochastic processes, and statistics}
\pacs{89.75.Da}{Systems obeying scaling laws}

\abstract{A recent exponent inequality is applied to a number of dynamical growth models. Many of the known exponents for models such as the Kardar-Parisi-Zhang (KPZ) equation are shown to be consistent with the inequality. In some cases, such as the Molecular Beam Equation, the situation is more interesting, where the exponents saturate the inequality. As the acid test for the relative strength of four popular approximation schemes we apply the inequality to the exponents obtained for two Non Local KPZ systems. We find that all methods but one, the Self Consistent Expansion, violate the inequality in some regions of parameter space. To further demonstrate the usefulness of the inequality, we apply it to a specific model, which belongs to a family of models in which the inequality becomes an equality. We thus show that the inequality can easily yield results, which otherwise have to rely either on approximations or general beliefs.}
\begin{document}
\maketitle

Growing interfaces is an important topic in Statistical Physics since the mid 80's, booming after the seminal works of Edwards and Wilkinson (EW) \cite{EW} and Kardar, Parisi and Zhang (KPZ) \cite{KPZ}. Since then, there has been great interest, both theoretical and experimental in such systems. This success is partially related to the fact that these systems exhibit generic properties of dynamical systems out of equilibrium, such as dynamical steady states, and dynamical phase transitions \cite{barabasi95,Halpin95}.

Nevertheless, in spite of the tremendous progress, except for a number of one dimensional exactly soluble problems \cite{EW,KPZ,Nonlocal,NMBE,FKPZ} and one two-dimensional problem \cite{AKPZ}, the sets of exponents given in the literature for many systems belonging to this class, vary considerably from author to author and depend strongly on the method of derivation \cite{KPZ,barabasi95,Halpin95,Hen91,SE,BC,Doherty,Katzav99,Moore01,Canet10}. Interestingly, simulations suffer from a similar problem, namely a large variety of results, which do not agree with each other - and even less agreement is found between them and the theoretical predictions \cite{simulations}. Researchers do not even agree as to whether an upper critical dimension exists or not \cite{BC,Moore01,Canet10,UCD}. This is very different from the situation in equilibrium phase transitions, where methods as different as high temperature expansion, momentum space RG and real space RG yield very close exponents \cite{HH}, and those being in agreement with simulations as well as experimental results. Under such circumstances rigorous results that can put bounds on the exponents describing the system are obviously most valuable.

Recently, a general response-correlation inequality in dynamical systems has been derived \cite{general}. This inequality is an extension of the Schwartz-Soffer inequality developed originally for quenched random problems \cite{SS}. The inequality is of a generic nature and relates the response at steady state of some measurable physical field to an external disturbance, to the time dependent correlations of that physical field. In the following, we explore the consequences of this inequality in various growth models. We begin with a short summary of the inequality, followed by a discussion of various analytical results that violate the inequality. This shows that the inequality is strong enough to point out problems in various approximation schemes, which may be the first step leading either to improving or discarding those schemes. Last, we show that in certain cases the inequality implies an equality, and demonstrate its potential usefulness.

Many interesting growth models, and actually many other dynamical systems can be described in terms of some physical field, $h(\mathbf{r},t)$ driven by a "noise" field, $\eta (\mathbf{r},t)$. This very broad setup includes any of the Langevin field equations, such as growth models of the KPZ family \cite{KPZ} and its many variants \cite{barabasi95,Halpin95}, noise driven Navier-Stokes \cite{McComb,ES02} etc... but even more complicated cases where such an equation cannot be explicitly written. In the context of growth models, the physical field $h(\mathbf{r},t)$ is simply a height function. Strictly speaking, $h(\mathbf{r},t)$ depends not only on the noise field at earlier times but also on initial conditions. The dependence on initial conditions decays, however, in time, and we are left with an implicit relation between the Fourier transform of the field and the Fourier transform of the noise $h(\mathbf{q},\omega )=h\{\mathbf{q},\omega ;\eta (\mathbf{l},\sigma )\}$, where $\eta (\mathbf{l},\sigma )$ is a Gaussian random field with zero mean $\left\langle \eta (\mathbf{l},\sigma ) \right\rangle =0$, and

\begin{equation}
\left\langle \eta (\mathbf{l},\sigma )\eta (\mathbf{m},\varsigma ) \right\rangle =2D_0 (l,\sigma )\delta (\mathbf{l}+\mathbf{m})\delta (\sigma +\varsigma )
\label{eq:1} \, .
\end{equation}
A comprehensive description of a physical system can be achieved in terms of the linear response function, which captures the response of the system to an external perturbation, and the $2$-point correlation function, which captures the spatio-temporal structure of the physical field. The response function, $G\left( q,\omega  \right)$, is defined by
\begin{equation}
\left\langle \frac{\delta h(\mathbf{q},\omega )}{\delta \eta (\mathbf{p},\sigma )} \right\rangle \equiv G(q,\omega )\delta (\mathbf{q}-\mathbf{p})\delta (\omega -\sigma )
\label{eq:2} \, ,
\end{equation}
and the $2$-point correlation function, $\Phi \left( q,\omega  \right)$, is defined by
\begin{equation}
\left\langle h(\mathbf{q},\omega )h(-\mathbf{p},-\sigma ) \right\rangle \equiv \Phi (q,\omega )\delta (\mathbf{q}-\mathbf{p})\delta (\omega -\sigma )
\label{eq:3} \, .
\end{equation}
In Ref.~\cite{general} the following exact inequality was proven
\begin{equation}
2{{\left| G(q,\omega ) \right|}^{2}}{{D}_{0}}(q,\omega )\le \Phi (q,\omega )
\label{eq:4} \, .
\end{equation}
This general inequality can be turned into a powerful exponent inequality. To achieve that let us recall the standard dynamic scaling picture in surface growth (see for example \cite{barabasi95,Halpin95,Vicsek84}) and the corresponding exponents.
First, the equal time $2$-point correlation function
\begin{equation}
\Lambda (q)=\Phi (q,t=0)=\int\limits_{-\infty }^{\infty }{d\omega }\Phi (q,\omega )
\label{eq:5} \, ,
\end{equation}
describes the static properties of the growing interface. It generally holds that $\Lambda (q)$
behaves as a power law in $q$ for small $q$, namely $\Lambda (q) \propto q^{-\Gamma}$. In the context of growth models $\Gamma =d+2\alpha $ where $d$ is the substrate dimension, and $\alpha$ is the roughness exponent \cite{barabasi95}. The larger the roughness exponent $\alpha$, the rougher the interface is.

The characteristic frequency, $\omega_C(q)$, associated with the decay in time of the correlation is given by \cite{SE,HH}
\begin{equation}
\omega_C^{-1}(q)=\pi \Phi (q,t=0)/\Lambda (q) \propto q^{-z}
\label{eq:7} \, ,
\end{equation}
where $z$ is known as the dynamic exponent. $\omega_C(q)$ has units of inverse time, and therefore, describes a characteristic equilibration time $t_X$ of the system as a function of its size $L$, namely $t_X \propto L^z$. Therefore, the larger the dynamic exponent, the larger the relaxation time is, and the slower the dynamics.

Another characteristic frequency, which can in general be different from $\omega_C(q)$, can be obtained from the response function, i.e. $\omega_R^{-1}(q) = G(q,0)$, which for small $q$ behaves like
\begin{equation}
\omega_R^{-1}(q) = G(q,0)\propto q^{-\bar{z}}
\label{eq:6} \, .
\end{equation}
The new exponent $\bar{z}$ describes the relaxation time of the system in response to an external perturbation.

Last, in this work we will focus our attention on the systems driven by spatially (but not temporally) correlated noise. In other words, we will consider the family of bare spectral functions $D_0(q,\omega)$ (i.e., the noise correlators in Eq.~(\ref{eq:1})) that for small $q$ and $\omega$ have the form
\begin{equation}
D_0(q,\omega) = B q^{-2\sigma} \quad \tx{with } \, \sigma \ge 0
\label{eq:8} \, ,
\end{equation}

The required exponent inequality is obtained now by setting $\omega =0$ in Eq.~(\ref{eq:4}),
\begin{equation}
2\bar{z}+2\sigma \le \Gamma +z
\label{eq:9} \, .
\end{equation}

The equation above relates three independent exponents. It turns out, however, that {\bf most} of the stochastic dynamical systems studied in the literature belong to one of two, not necessarily mutually exclusive, classes \cite{general}. In each of those classes the exponent $\bar{z}$ obeys a different scaling relation, relating it to the other exponents such as $\Gamma ,z$ and $\sigma$. (Note that we do not claim that {\bf all} stochastic dynamical systems belong to one of the two, e.g. the stochastic Sine-Gordon systems \cite{SG,NSG}).

Class $\tx{I}$ is that of generalized Hamiltonian systems where $\bar{z}=\Gamma -2\sigma$. This is the case for example in Model A of Hohenberg and Halperin \cite{HH} (i.e., the celebrated $\phi^4$ theory) where $\bar{z}=\Gamma$. Class $\tx{II}$ is the class of Galilean invariant systems. It is composed of systems whose dynamics is given by a Langevin field equation of the form
\begin{equation}
\gamma \frac{\partial h(\mathbf{q})}{\partial t}=F_{\mathbf{q}}\{h\}+\eta (\mathbf{q},t)
\label{eq:10} \, ,
\end{equation}
supplemented with the condition ${\delta {{F}_{\mathbf{q}}}\{h\}}/{\delta h\left( \mathbf{q}=0 \right)}\;=0$. For growth models of the KPZ family \cite{barabasi95} this corresponds to the fact that the dynamics is invariant to translations in the h-direction, i.e. with respect to $h \to h + \tx{Const}$, also known as the Galilean Invariance. For this class it was shown in Ref.~\cite{general} that $z=\bar{z}$. It is important to mention that the proof is based on the Family-Vicsek dynamical scaling \cite{barabasi95,BC,Doherty,Moore01,Vicsek84}
\begin{equation}
G(q,\omega )=\frac{1}{{q^{\bar{z}}}}f\left( \frac{\omega}{\omega_q} \right)
\label{eq:11} \, ,
\end{equation}
Note, that most of the growth models belong to Class $\tx{II}$, including the KPZ equation \cite{KPZ}, the Molecular Beam Equation (MBE) \cite{Lai} and many more \cite{barabasi95,Halpin95}. Actually the noise driven Navier-Stokes equation \cite{McComb,ES02}, various wetting-front models \cite{Golestanian,LeDoussal,Katzav07} and crack propagation equations \cite{RF,Katzav06} also belong to this class.

Using the relation $\bar{z}=z$, the general inequality (\ref{eq:9}) reduces to the simpler
\begin{equation}
z\le \Gamma -2\sigma \le \Gamma
\label{eq:12} \, .
\end{equation}
The above inequality is correct for any set of exponents, whether strong coupling or weak coupling. In fact, for weak coupling exponents, we expect the leftmost inequality to hold as an equality.

The most famous systems belonging to Class $\tx{II}$ are growth models of the KPZ \cite{KPZ,barabasi95} and MBE families \cite{barabasi95,Villain90,Villain91,Lai}. In those systems, we can obtain more than just an inequality relating the dynamic exponent $z$ and the steady state exponent $\Gamma $. This is because the two are related by a scaling relation $(\Gamma -d)/2+z=2$ for the KPZ family and $(\Gamma -d)/2+z=4$ for the MBE family \cite{barabasi95,Lai}. (Recall that $\alpha =(\Gamma -d)/2$ is the roughness exponent). It is easy to verify that all the known results for regular KPZ obey the inequality, which becomes in this case $z \le (d+4)/3$. For $d=1$ all the analytic methods recover the exact results for regular KPZ, namely $z=3/2$ that obeys the inequality $z \le 5/3$. Therefore, there is no surprise that in one dimension the inequality is obeyed for regular KPZ by all the methods. For higher dimensions, although the exponents obtained analytically \cite{KPZ,barabasi95,Halpin95,Hen91,SE,BC,Doherty,Katzav99,Moore01,Canet10} deviate most considerably from the simulations \cite{simulations} all methods yield $\Gamma >d$ and $z<2$, and so the inequality becomes less tight.

For the MBE equation \cite{Lai}, the inequality~(\ref{eq:12}) becomes $z \le (d+8)/3$. The one-loop Dynamic Renormalization Group (DRG) result \cite{Lai}, as well as the Self Consistent Expansion (SCE) \cite{MBE02} yield $z = (d+8)/3$, and so the inequality is saturated.

To demonstrate the usefulness of the exponent inequality, we use it to test various theoretical approaches to the study of the KPZ family, and concentrate on two non-local extensions to the KPZ equation. From the early days of the field it was clear that the simple models such as the EW and KPZ models show systematic deviations from experimental data. However, since the general scaling picture developed in this context \cite{barabasi95,Halpin95,Vicsek84} usually applies, various models with correlated noise \cite{TCN} conserved \cite{Villain90,Villain91,Lai}, nonlocal \cite{Mukh97,chat99,Jung98} and fractional dynamics \cite{FKPZ,Mann01} have been developed to account for important physical effects that were not taken into account beforehand.

A Non-local KPZ (NKPZ) equation has been introduced in \cite{Mukh97} to account for the non-local hydrodynamic interactions in deposition of colloidal particles in a fluid. It was later generalized to spatially correlated noise ($\sigma \ne 0$) in Ref.~\cite{chat99}. The non-local KPZ equation suggested in \cite{Mukh97} for the height function, $h(\mathbf{r},t)$ of the deposited material is given by
\vspace{1cm}
\begin{eqnarray}
&&\frac{\partial h\left( \mathbf{r},t \right)}{\partial t}=\nu \nabla^2 h\left( \mathbf{r},t \right)\\
&&+\int{d\mathbf{r'}g\left( {\mathbf{r'}} \right)\nabla h\left( \mathbf{r}+\mathbf{r'},t \right)\cdot \nabla h\left( \mathbf{r}-\mathbf{r'},t \right)}+\eta \left( \mathbf{r},t \right) \nonumber
\label{eq:13} \, ,
\end{eqnarray}
where the kernel $g\left( \mathbf{r} \right)$ has a short range part, $\lambda_0 \delta \left( \mathbf{r} \right)$, and a long-range part $\sim \lambda_\rho r^{\rho -d}$. In Fourier space, $\hat{g}\left( q \right)={{\lambda }_{0}}+\lambda_\rho q^{-\rho}$ ($\rho$ is the non-locality parameter). For simplicity, we discuss here only the case $\lambda_0 = 0$. The noise has zero mean, but is allowed to have general correlations of the form given by equation (\ref{eq:8}). The strong coupling solution found by the DRG is \cite{Mukh97,chat99}
\begin{equation}
z_{DRG}=2+\frac{\left( d-2-2\rho  \right)\left( d-2-3\rho  \right)}{\left( 3+{{2}^{-\rho }} \right)d-6-9\rho }
\label{eq:14} \, .
\end{equation}
Because of the extra scaling relation between $\alpha$ and $z$ \cite{Mukh97,chat99}, namely $\alpha +z=2-\rho $, there is only one independent exponent in NKPZ.

\begin{figure}[ht]
\centerline{\includegraphics[width=8cm]{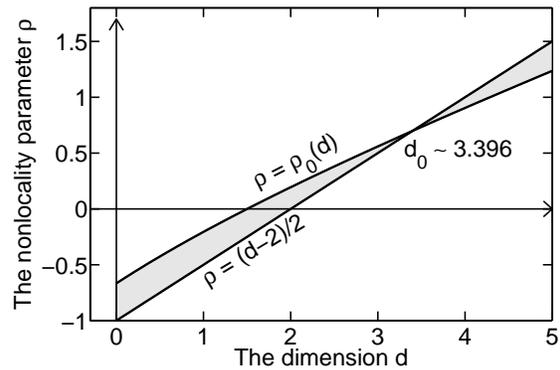}}
\caption{Violation of the response-correlation inequality (\ref{eq:12}) with $\sigma =0$ by the DRG method derived in Refs.~\cite{Mukh97,chat99} occurs in the shaded area enclosed by the curves $\rho =(d-2)/2$ and $\rho =\rho_0(d)$ in the phase diagram.}
 \label{fig:DRG}
\end{figure}

The above result violates the inequality (\ref{eq:12}) over a whole range of parameters defined by $\Gamma _{DRG}-z_{DRG}-2\sigma <0$. To be more concrete, let us specialize to the case $\sigma =0$ where the inequality, $\Gamma_{DRG} - z_{DRG}<0$ is violated in
\begin{equation}
\frac{d-2}{2}\le \rho \le \rho_0 (d) \quad \tx{for} \quad d < d_0
\label{eq:15} \, ,
\end{equation}
and
\begin{equation}
\rho_0 (d) \le \rho \le \frac{d-2}{2} \quad \tx{for} \quad d>d_0
\label{eq:16} \, ,
\end{equation}
where $\rho _0(d)=\frac{d-2}{3}+\frac{W\left( \frac{1}{9}{2^{\frac{2-d}{3}}}d\ln 2 \right)}{\ln 2}$, $d_0 \simeq 3.395$ is the solution of $\rho_0 (d_0)=\frac{d_0-2}{2}$ and $W\left( x \right)$ is the Lambert function. The shaded region in Fig.~\ref{fig:DRG} is the region in the $(d,\rho)$ plane where the inequality (i.e. Eq.~(\ref{eq:12}) with $\sigma =0$) is violated.

\begin{figure}[ht]
\centerline{\includegraphics[width=8cm]{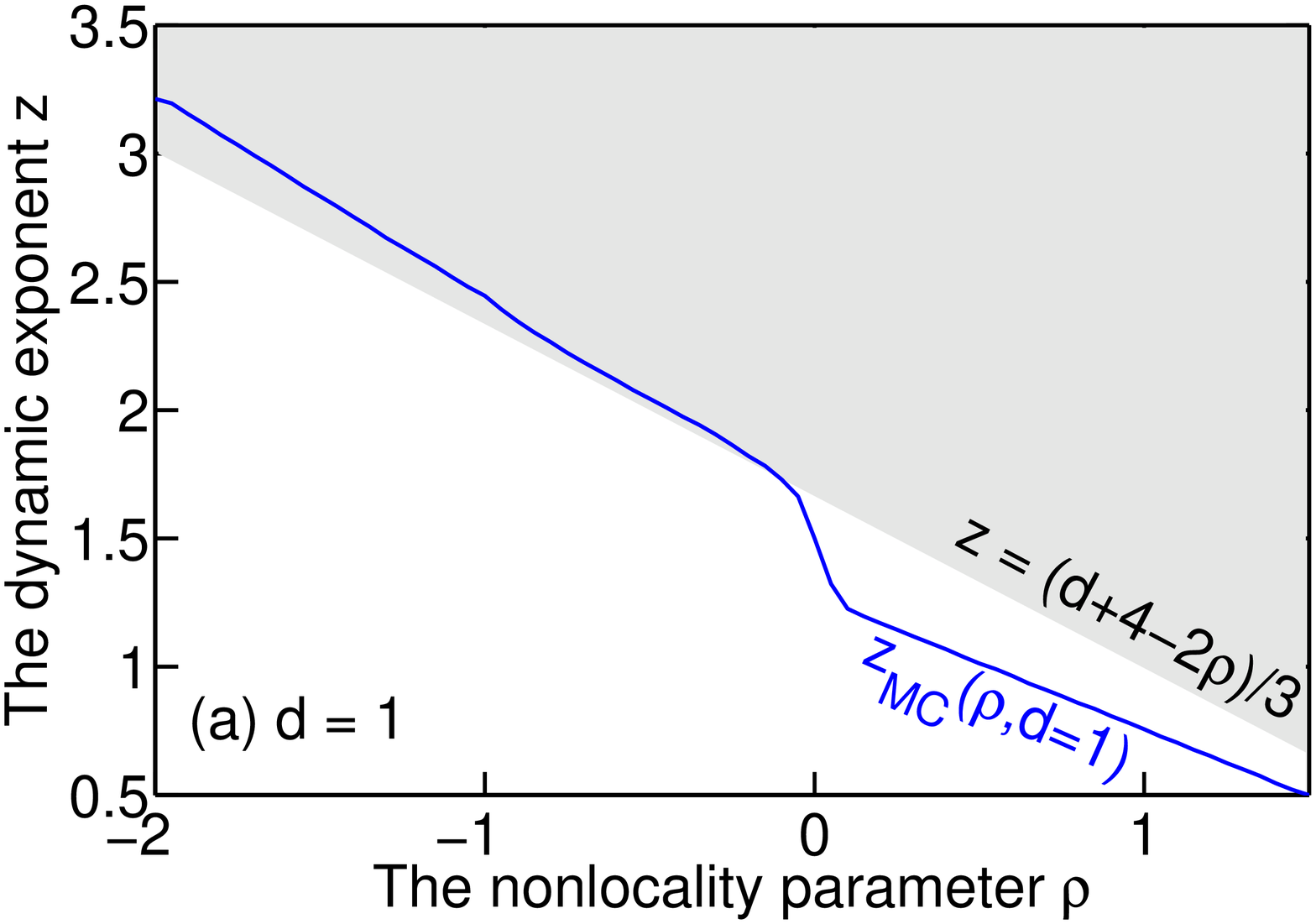}}
\centerline{\includegraphics[width=8cm]{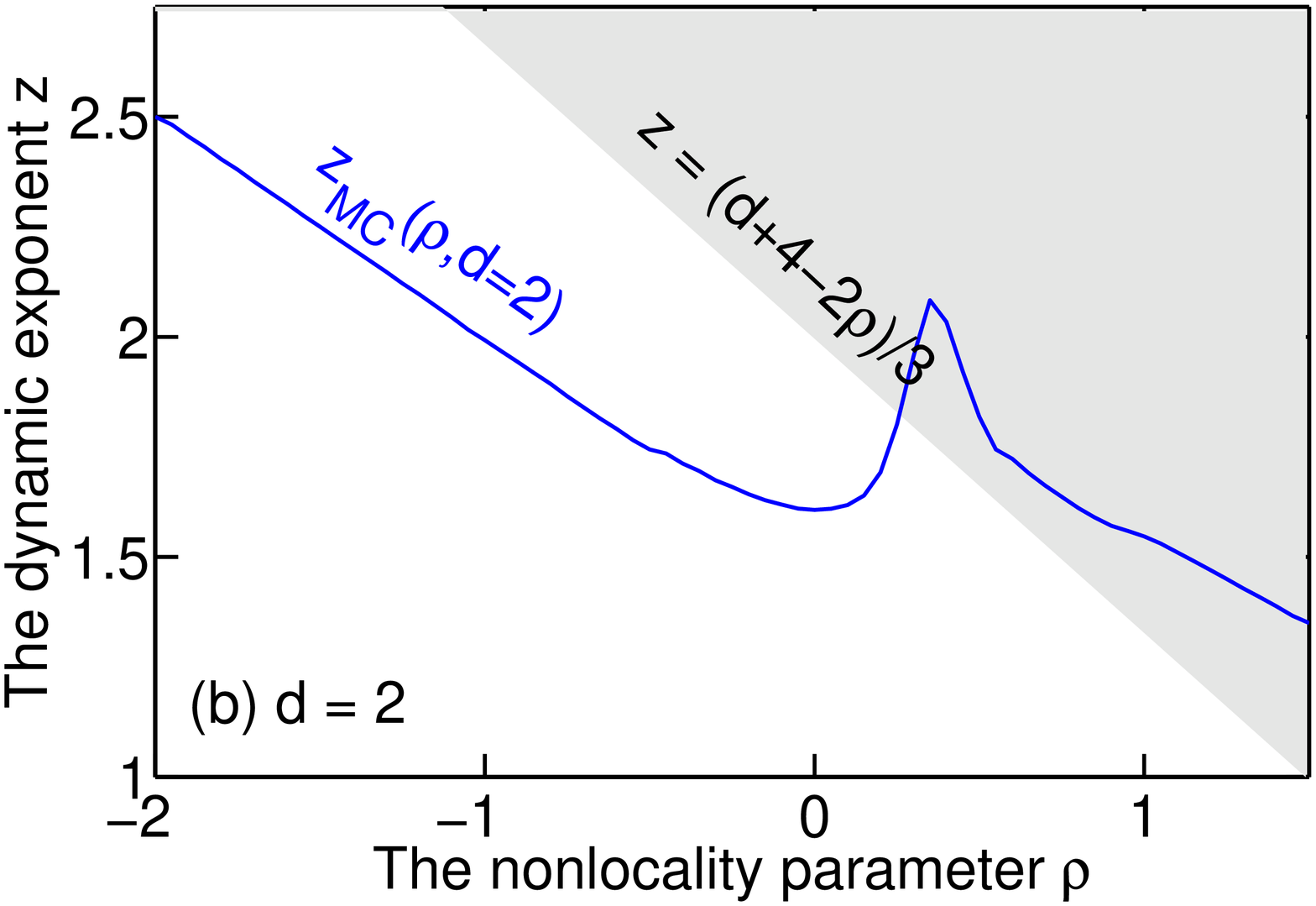}}
\centerline{\includegraphics[width=8cm]{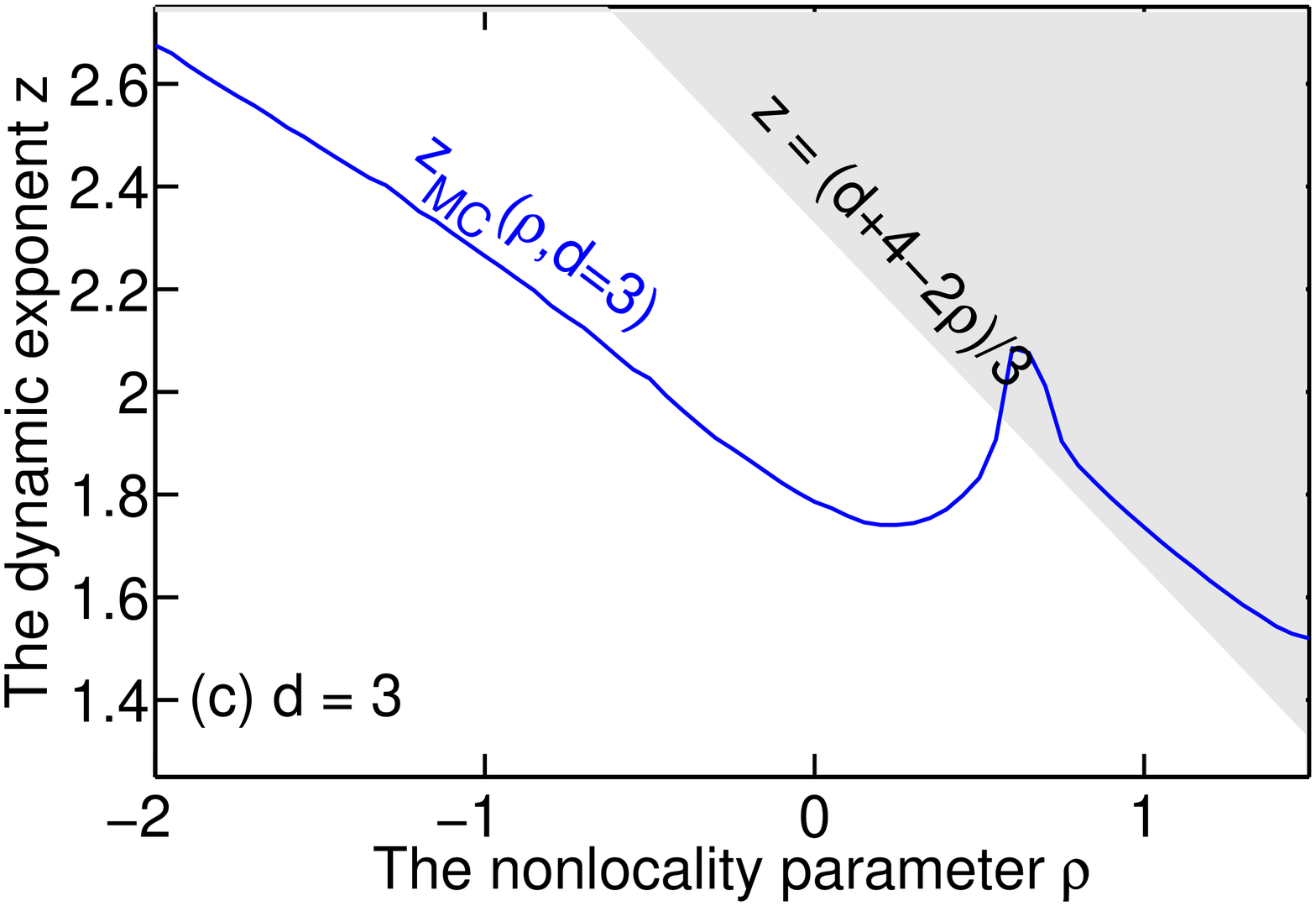}}
\caption{(Colour online) The Dynamic exponent $z$ as a function of the non-locality parameter $\rho$ using the mode coupling result (solid (blue) line - reproduced from Ref.~\cite{Hu02}). The part of the solid (blue) line within the shaded region is excluded by the inequality (\ref{eq:12}) with $\sigma =0$. As can be seen the mode coupling result violates the inequality in certain regions in all dimensions.}
 \label{fig:MC}
\end{figure}

We now turn to results derived for the same NKPZ model (\ref{eq:13}) using an improved Mode-Coupling (MC) approach originally derived for the local KPZ problem in \cite{Moore01} and later applied by Hu and Tang \cite{Hu02} to the non-local case. In Fig.~\ref{fig:MC} we give the solution for the dynamic exponent $z$ for $d=1,2,3$ reported in \cite{Hu02}. The shaded area bounded by the line $z=\left( d+4-2\rho  \right)/3$ marks the region where the inequality is violated by the MC prediction.

\begin{figure}
\centerline{\includegraphics[width=8cm]{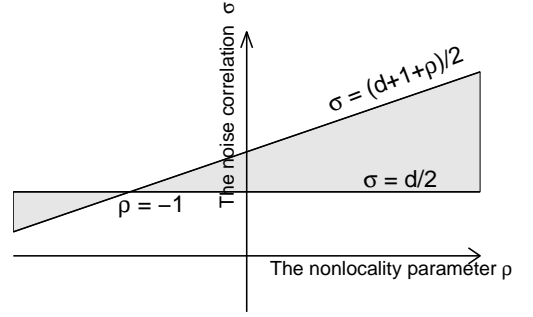}}
\caption{Violation of the response-correlation inequality (\ref{eq:12}) by the SA method derived in Ref.~\cite{Tang01} occurs in the shaded area enclosed by the curves $\rho =-1$ and $\sigma ={\left( d+1+\rho  \right)}/2$ in the phase diagram.}
 \label{fig:SA}
\end{figure}

Next, we check for violations of the inequality in results obtained by Tang and Ma \cite{Tang01} generalizing the Flory–-type Scaling Approach (SA) of Hentschel and Family \cite{Hen91} to the non-local case. The strong-coupling dynamical exponent obtained by using this method is
\begin{equation}
z_{SA}=\frac{\left( 2-\rho  \right)\left( d+2-2\sigma  \right)}{d+3-2\sigma}
\label{eq:17} \, .
\end{equation}
It turns out that this solution violates the inequality in a whole region of parameter space defined by
\begin{equation}
\frac{d}{2}<\sigma <\frac{d+1+\rho }{2}\quad \tx{for} \quad \rho >-1
\label{eq:18} \, ,
\end{equation}
and
\begin{equation}
\frac{d+1+\rho }{2}<\sigma <\frac{d}{2}\quad \tx{for} \quad \rho <-1
\label{eq:19} \, .
\end{equation}
These results are presented graphically in Fig.~\ref{fig:SA}.

Last we tested the results for NKPZ of the Schwartz-–Edwards Self-Consistent Expansion (SCE) \cite{SE,Katzav99}. This method predicts a whole zoo of possible phases all of which are consistent with the inequality, and therefore the only method so far that does not contradict the inequality.

\begin{figure}
\centerline{\includegraphics[width=8cm]{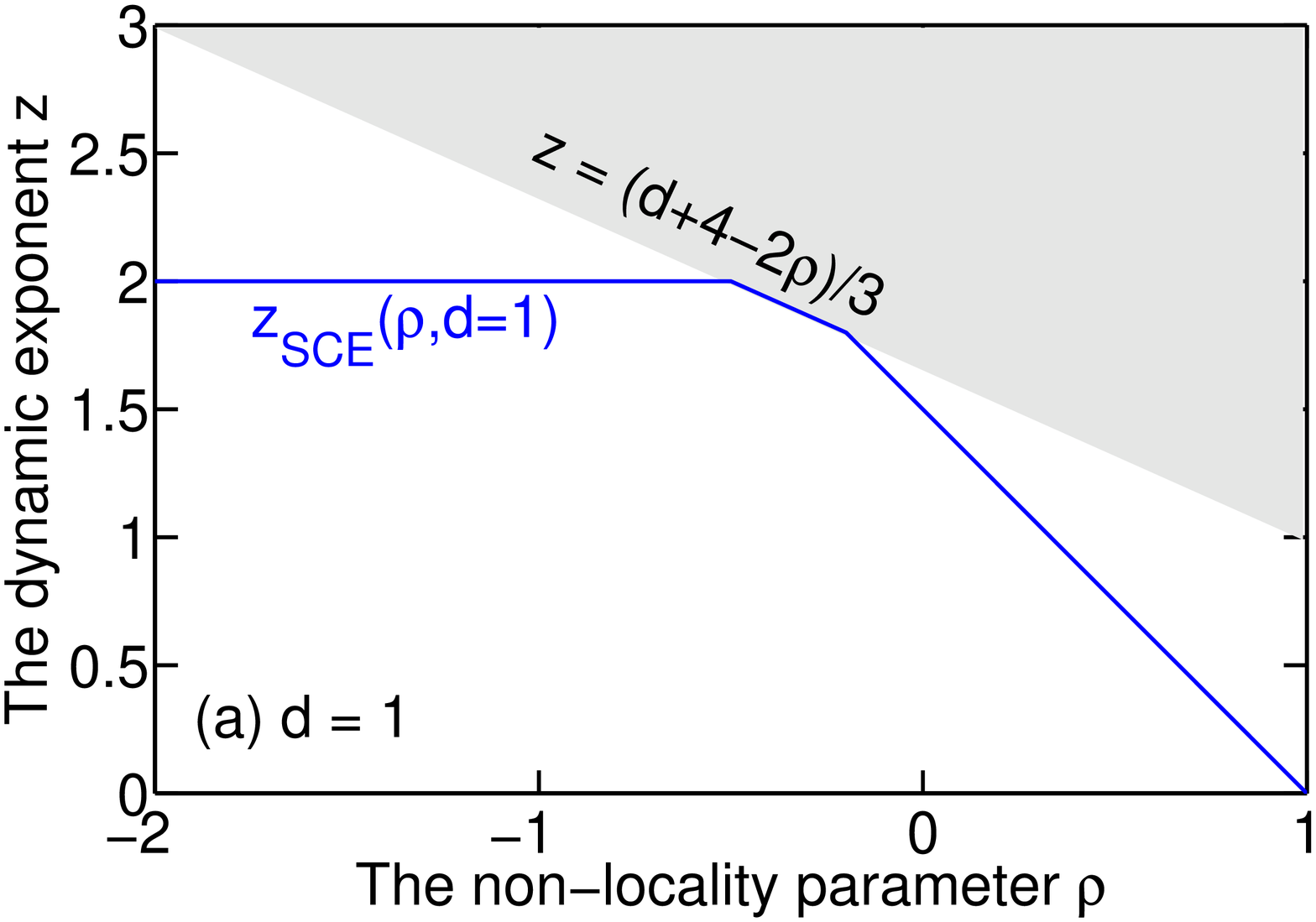}}
\centerline{\includegraphics[width=8cm]{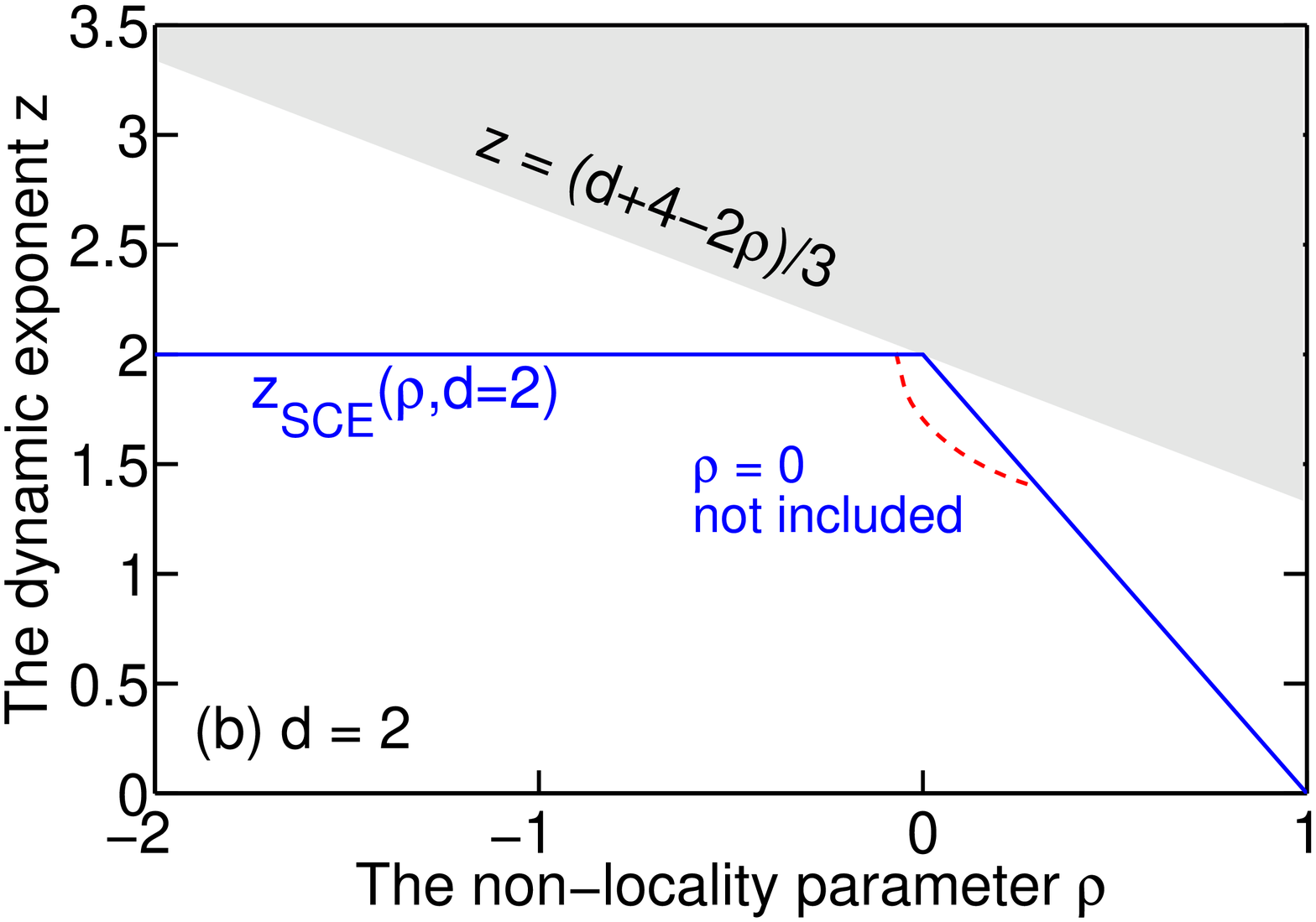}}
\centerline{\includegraphics[width=8cm]{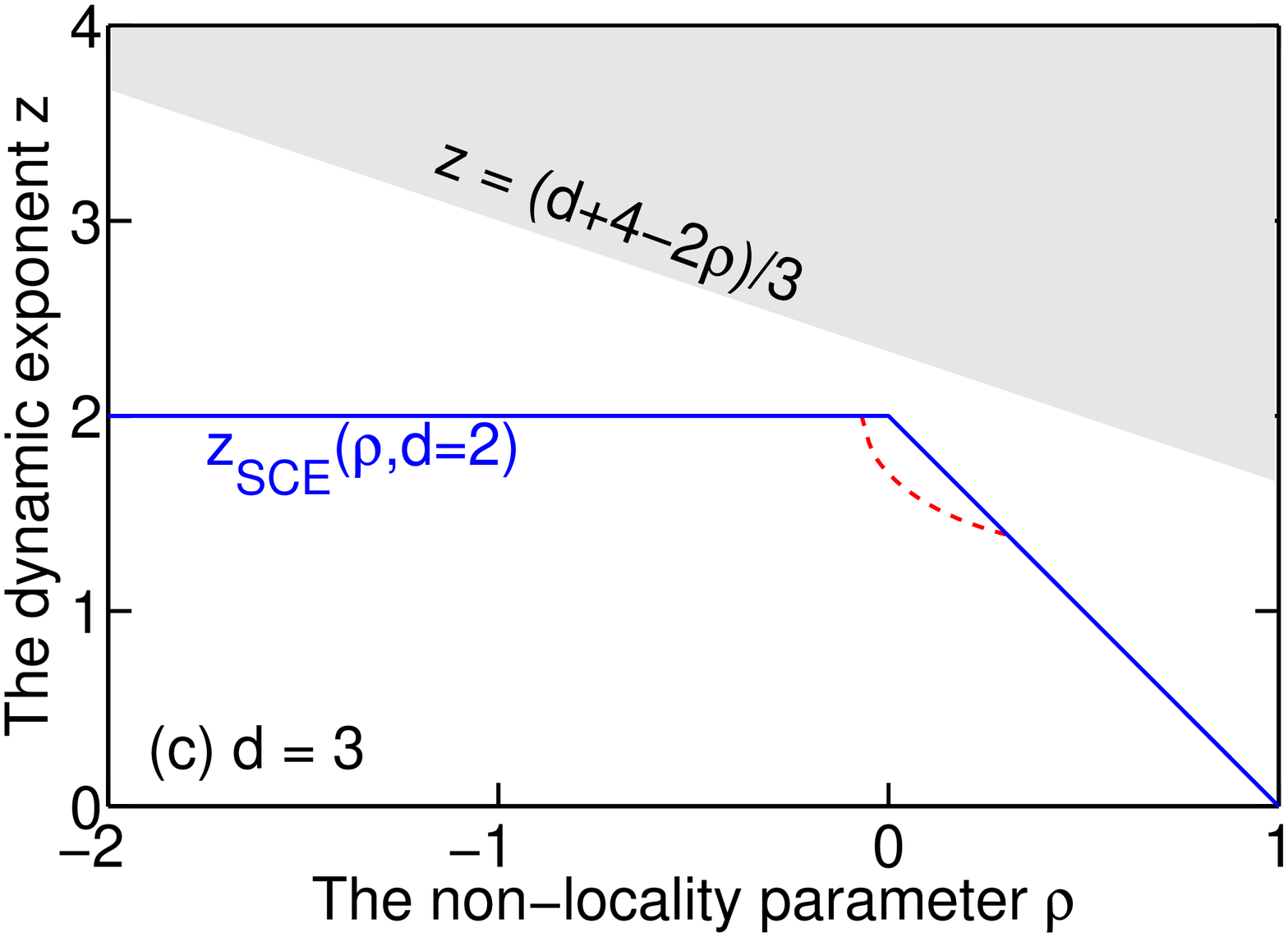}}
\caption{(Colour online) – The dynamic exponent $z$ as a function of the non-locality parameter $\rho$ for uncorrelated noise $(\sigma =0)$ in $d=1,2,3$ dimensions resulting from the Self-Consistent Expansion \cite{NKPZ03}. For $d=1$ the solution saturates the inequality for $-{1}/{2}\;<\rho <-{1}/{5}\;$. Note that for $d=2,3$ two phases are possible in some range of the parameters as seen from the fact that $z$ is a multivalued function of $\rho$ (solid (blue) line – weak coupling solution, and dashed (red) line – strong coupling solution). Also note that for $d=2$ and $\rho=0$ there is only one phase as the solid (blue) line does not exist for $\rho =0$.}
 \label{fig:SCE}
\end{figure}

To check if the picture we get so far may be more general we consider another non local system, which has been introduced and analyzed in Ref.~\cite{NKPZ03}. In this model, the nonlinearity in Eq.~(\ref{eq:13}) is replaced by $\int d\mathbf{r'} \frac{\nabla h\left( \mathbf{r} \right) \cdot \nabla h\left( \mathbf{r}-\mathbf{r'} \right)}{|\mathbf{r}-\mathbf{r'}|^{d-\rho}}$, such that the contribution to the growth at each point comes from the interaction of the gradient at that point with all the other points on the interface, rather than from all pairs at equal distance. The Scaling-Approach mentioned above yields the same results as before for this model as it depends only on the dimensions of the various quantities involved, and ignores the exact spatial structure, and therefore, gives rise to violations of the inequality here too.
The situation with the DRG approach is even more severe as the theory generates more relevant terms under renormalization, and therefore not surprisingly produces inconsistencies with the inequality.
Interestingly, the results of the SCE method (reported in Ref.~\cite{NKPZ03}) are not only consistent with the exact one-dimensional result \cite{Nonlocal} but the exponents in all the phases are consistent with our stronger, $\sigma $-dependent inequality over the whole parameter space and all dimensionalities. In Fig.~\ref{fig:SCE} we give the SCE solution for the dynamic exponent $z$ for $d=1,2,3$ for uncorrelated noise ($\sigma =0)$(but as stated above the inequality holds also for $\sigma >0$).

The inequality is shown to be a useful tool in detecting shortcomings of various analytical methods applied to the NKPZ family. This suggests that results obtained by those  methods should be suspected even in cases where the inequality is not violated ,at least until the origin of violation of the inequality is understood.

The last demonstration of the usefulness of the inequality, is when a system belongs to both Class $\tx{I}$ and Class  $\tx{II}$, i.e. when the system is Hamiltonian as well as Galilean Invariant. In these cases the inequality becomes an equality \cite{general}, namely $z=\Gamma-2\sigma$. This reduces the number of unknown exponents by one, and can be a powerful tool when dealing with certain nonlinear growth models. For example, a surface evolving due to deposition in the presence of surface tension can be described by the Hamiltonian $H=\frac{\gamma}{2} \int{{{d}^{d}}\mathbf{r}\sqrt{1+{{\left( \nabla h \right)}^{2}}}}$, which leads to the following equation
 \begin{equation}
{{\partial }_{t}}h\left( \mathbf{r},t \right)=\gamma \nabla \cdot \left( \frac{\nabla h}{\sqrt{1+{{\left( \nabla h \right)}^{2}}}} \right)+\eta \left( \mathbf{r},t \right)
\label{eq:20} \, .
\end{equation}
This model is believed to belong to the EW universality class \cite{barabasi95} on the basis of symmetry arguments, or by using a small-gradient expansion that leads directly to the EW equation \cite{EW}. The inequality shows rigorously that $z=d+2\alpha$, avoiding issues of the validity of the small-gradient expansion. This actually allows deriving quick and accurate results when applicable.

To summarize, in this paper we have shown how to use a recent inequality derived in \cite{general} to growth models described in terms of stochastic field theories. We show that the inequality, which involves the correlation and the response functions can be translated into a simple inequality for the scaling exponents $\Gamma$, $z$ and the noise correlation exponent $\sigma$, in cases where a single $q$ dependent time scale exists. Although being extremely simple, this inequality can be quite powerful when examining analytical, numerical and experimental results.

To demonstrate the utility of the inequality, we reviewed analytical results for two non-local KPZ models, obtained by using four different methods: Dynamical Renormalization Group, Mode-Coupling, Scaling-Approach and the Self-Consistent Expansion. Interestingly, the first three methods yield results which contradict the inequality for a whole set of parameters, while the Self-Consistent Expansion is the only one which never violates it. This has an important implication on the choice of analytical tools when dealing with such stochastic models. Also, we have pointed out the observation that whenever the system is Hamiltonian as well as Galilean invariant, the general inequality implies an equality $z=\Gamma-2\sigma$, and can be quite powerful when applicable. Last, even in cases where the inequality is not saturated nor implies an exact equality, in effect it can provide a reasonable estimate for the scaling exponents. In the cases we have reviewed, the bounds implied by the inequality do not deviate a lot from the real result, especially in low dimensions. This suggests using the inequality to derive quick approximate estimates for nonlinear systems.

This work has interesting experimental implications. If in a given experimental system $z\ne \bar{z}$ we get a strong constraint on the kind of theory that could describe the phenomenon under consideration, namely a theory that cannot have Galilean invariance. Similarly, if experimentally $\Gamma \ne \bar{z}$ the theory cannot be Hamiltonian. This can provide useful guidelines for both theorists and experimentalists who may wish to model the system.

There are obviously some interesting open questions. For example, are there relevant growth models that violate the Galilean invariance, and therefore have distinct $z$ and $\bar{z}$? More generally, the question here is regarding the existence of interesting classes other then the two identified so far. Another direction could be to explore similar inequalities for systems described by more than one-field \cite{many}, and systems with quenched disorder \cite{quenched}.

\end{document}